\documentclass[aps,showpacs,twocolumn,amsmath,amssymb,prl]{revtex4-1} 

\usepackage{graphicx} 
\usepackage{bm}

\begin{document} 
 
\title{Nonlinearly-enhanced energy transport in many dimensional quantum chaos} 
 
\author{D. S. Brambila$^{1,2}$ and A. Fratalocchi$^1$} 

\email{andrea.fratalocchi@kaust.edu.sa} 
\homepage{www.primalight.org}

\affiliation{
$^1$PRIMALIGHT, Faculty of Electrical Engineering; Applied Mathematics and Computational Science, King Abdullah University of Science and Technology (KAUST), Thuwal 23955-6900, Saudi Arabia\\
$^2$Max-Born Institute, Max-Born-Stra$\beta$e 2 A, 12489, Berlin, Germany
}


\date{\today} 
 
\begin{abstract} 
By employing a nonlinear quantum kicked rotor model, we investigate the transport of energy in multidimensional quantum chaos. Parallel numerical simulations and analytic theory demonstrate that the interplay between nonlinearity and Anderson localization establishes a perfectly classical correspondence in the system, neglecting any quantum time reversal. The resulting dynamics exhibits a nonlinearly-induced, enhanced transport of energy through soliton wave particles.      
\end{abstract} 
 
\pacs{05.45.Mt, 05.45.Yv, 03.75.-b} 
 
\maketitle 
Anderson localization is a fundamental concept that, originally introduced in solid-state physics to describe conduction-insulator transitions in disordered crystals, has permeated several research areas and has become the subject of great research interest \cite{sheng,scheffold99:_local_or_class_diffus_of_light,schwartz07:_trans_and_ander_local_in,PhysRevLett.96.043902,conti08:_dynam_light_diffus_three_dimen,RevModPhys.57.287,PhysRevLett.100.013906,billy08:_direc_obser_of_ander_local,roati08:_ander_local_of_non_inter,efetov97:_super_in_disor_and_chaos,PhysRevLett.49.509}. Theories and subsequent experiments demonstrated that disorder favors the formation of spatially localized states, which sustain diffusion breakdown and exponentially attenuated transmission in random media \cite{sheng}. Although many properties of wave localization are now well understood, several fundamental questions remains. Perhaps one of the most intriguing problem is related to the transport of energy. Intuitively, one can expect that disorder ---by favoring exponentially localized stated--- arrests in general any propagation inside a noncrystalline medium. However, the interplay between localization and disorder is nontrivial \cite{conti08:_dynam_light_diffus_three_dimen,Molinari:12} and under specific conditions randomness can significantly enhance energy transport. In particular, it has been observed that quasi-crystals with multifractal eigenstates and/or material systems with temporal fluctuations of the potential (or refractive index), lead to anomalous diffusion in the phase space \cite{PhysRevLett.70.3915,levi11:_disor_enhan_trans_in_photon_quasic,PhysRevLett.39.1424,PhysRevLett.82.4062}. This originates counterintuitive dynamics including ultralow conductivities \cite{PhysRevLett.70.3915}, as well as the formation of mobility edges even in one dimensional systems \cite{PhysRevLett.82.4062}. All these studies focused on specific geometries and linear materials, while nothing is practically known about the role of nonlinearity in enhancing (or depleting) the transport of energy in disordered media. This problem acquires a strong fundamental character when refereed to the field of quantum localization. In this area, quantum-classical correspondences mediated by Anderson localization possess many implications in the irreversible behavior of time reversible systems, which are at the basis of a long standing physical dispute ---i.e., the Loschmidt paradox \cite{Loschmidt}--- as well as many fascinating quantum phenomena such as the time reversal of classical irreversible systems and the quantum echo effect \cite{PhysRevLett.101.074102,PhysRevLett.61.659}. It has been argued, in particular, that microscopic chaos is at the basis of the irreversible entropy growth observed in classical systems \cite{gaspard98:_exper}. Time reversal, according to this interpretation, is only possible at the quantum level \cite{PhysRevLett.101.074102,PhysRevLett.61.659} and sustained by Anderson localization, which breaks diffusive transport and suppresses the mixing ability of chaos \cite{PhysRevA.44.R3423}. However, when more dimensions are considered, numerical simulations predict that ergodicity is fully restored and diffusive transport settles is again, thus re-establishing the classical features of chaos and preventing quantum time reversal \cite{PhysRevLett.61.659}. Nevertheless, theoretical work reported to date considered only noninteracting systems, characterized by linear equations of motion. The Loschmidt paradox, conversely, involved the use of interacting atoms, whose interplay in the mean field regime is accounted by short and/or long ranged nonlinear responses \cite{PhysRevLett.94.160401,PhysRevLett.100.240403,griffin95:_bose_einst_conden}. Besides that, as pointed out in the literature \cite{PhysRevLett.101.074102}, atoms interactions are of crucial importance in quantum localization and diffusion. A key question therefore lies in understanding the role of nonlinearity in transporting energy in multidimensional quantum chaos.\\
In this Letter, we theoretically investigate this problem by employing both numerical simulations and analytic techniques. To pursue a general theory, we here consider the following two dimensional model:
\begin{align}
\label{mod}
&i\frac{\partial\psi}{\partial t}+\nabla^2\psi+\int d\mathbf{r} R(\mathbf{r}'-\mathbf{r})\psi(\mathbf{r}')+U\psi\delta_T(t)=0,
\end{align}
with $\mathbf{r}=(x,y)$, $\nabla^2=\partial^2/\partial x^2+\partial^2/\partial y^2$, $\delta_T=\sum_n\delta(t-nT)$ a periodic delta-function of period $T$, $R$ a general nonlinear response and $U(x,y)=\gamma(\cos x+\cos y)+\epsilon\cos(x+y)$ a two dimensional periodic potential with strength defined by $\epsilon$ and $\gamma$. Equation (\ref{mod}) defines a two dimensional, nonlinear quantum kicked rotor: for $R=0$ it reduces to the linear quantum kicked rotator \cite{PhysRevLett.61.659} while for $U=0$ it corresponds to the 2D nonlinear Schr\"odinger equation (NLS), which represents a universal model of nonlinear waves in dispersive media \cite{griffin95:_bose_einst_conden}. In one dimension, conversely, Eq. (\ref{mod}) generalizes the nonlinear model investigated in \cite{PhysRevA.44.R3423} with classical chaos parameter $K=2\gamma T$. Despite its deterministic nature, Eq. (\ref{mod}) can be precisely mapped to the Anderson model with a random potential \cite{haake01:_quant_signat_chaos}, and therefore furnishes a fundamental model for studying energy transport in random systems. The nonlinear response $n=\int d\mathbf{r} R(\mathbf{r}'-\mathbf{r})\psi(\mathbf{r}')$ is modeled as a nonlocal term following a general diffusive nonlinearity $(1-\sigma^2\nabla^2)n=|\psi|^2$, with nonlocality controlled by $\sigma$. When $\sigma=0$, the system response is local with $n=|\psi|^2$. For $\sigma\neq 0$, conversely, the system nonlinearity becomes long ranged with kernel given by $R(\mathbf{r})=\frac{1}{2\pi}K_0(\frac{\mathbf{r}}{\sigma})$, being $K_0$ the modified Bessel function of second kind. Diffusive nonlinearities are particularly interesting in the context of nonlinear optics, as they can be easily accessed in liquids \cite{PhysRevLett.102.083902,fratalocchi:044101}, as well as in Bose-Einstein Condensates (BEC), where they generalize previously investigated models \cite{PhysRevLett.95.200404,Bang02}.\\
\begin{figure}
\centering
\includegraphics[width=8.5cm]{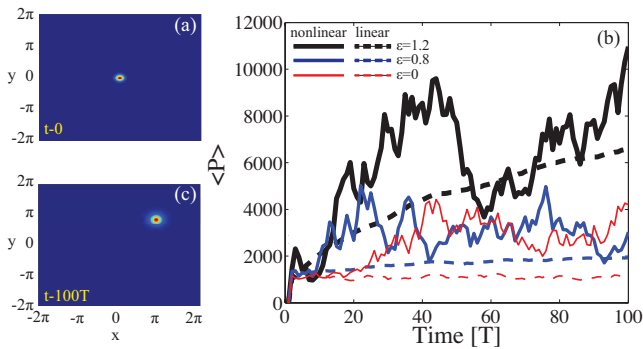}
\caption{
\label{sim}
(Color Online). (a)-(b) spatial density $|\psi|^2$ distribution at (a) $t=0$ and at (b) $t=100 T$;  
(c) momentum diffusion $\langle P\rangle$ versus time in linear (dashed lines) and nonlinear (solid lines) conditions and for increasing coupling $\epsilon$. In the simulations we set $\sigma=0.2$, $\omega_0=0.3$, $A=4$ and $K=1.8$.}
\end{figure}
We begin our theoretical analysis by calculating the momentum diffusion $\langle P\rangle=\langle\psi|\frac{\hat{p}^2}{2}|\psi\rangle$ versus time, with $\hat{p}=\nabla/i$ the momentum operator and $\langle\psi| f|\psi\rangle=\int d\mathbf{r}f|\psi|^2$ the quantum average. Parallel numerical simulations are performed by a direct solution of (\ref{mod}) with an unconditionally stable algorithm. In order for the field $\psi$ to explore the periodic potential $U$, we here consider wave packets whose spatial extension $\Delta r\ll 2\pi$. Figure \ref{sim} summarizes our results obtained for $\sigma=0.2$, by launching at the input a gaussian beam $\psi=Ae^{-x^2/\omega_0^2}$ with waist $\omega_0=0.3$ and amplitude $A=4$ (Fig. \ref{sim}a). The stochastic parameter $K$ has been set to $K=1.8>K^*$, above the stochastization threshold $K^*\approx 0.97$ where the linear classical uncoupled rotor exhibits diffusive transport in momentum space \cite{PhysRevLett.61.659}. For comparison, we also calculated the linear dynamics resulting from $R=0$ (Fig. \ref{sim}b dotted line). As seen from Fig. \ref{sim}b, the 2D nonlinear rotor behaves dramatically different with respect to its linear counterpart, demonstrating the strong role played by nonlinearity in the process. In particular, the linear system exhibits Anderson localization and diffusion suppression for $\epsilon=0$ (uncoupled condition), while for growing $\epsilon$ it shows a monotonically increasing sub-diffusion (Fig. \ref{sim}b). In the nonlinear regime, conversely, Anderson localization is suppressed even for $\epsilon=0$, and the dynamics shows an erratic, random-like behavior that does not manifest any simple monotonic increase for growing values of $\epsilon$. These results are also significantly different from the nonlinear kicked rotor in one dimension \cite{PhysRevA.44.R3423}, where nonlinearity was observed to induce localization effects.
\begin{figure}
\centering
\includegraphics[width=8.5cm]{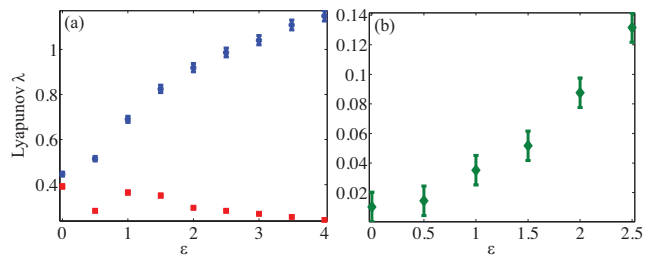}
\caption{
\label{lyap}
(Color Online). Positive Lyapunov exponent $\lambda$ versus coupling $\epsilon$, calculated for (a) Eqs. (\ref{sm0}) and (b) Eq. (\ref{wa1}). In the simulation we set $K=5$.}
\end{figure}
To theoretically understand this dynamics, we reduce the system to a nonlinear map modeling the nonlinear evolution of the ground state of Eq. (\ref{mod}). This analysis is justified by the observation that the spatial field profile, despite the chaotic motion, is not significantly altered in time (Fig. \ref{sim}a,c). Due to the nonintegrability of the 2D NLS equation, we found analytical expressions by a variational analysis \cite{PhysRevLett.95.200404,Bang02}. In particular, we begin from the Lagrangian density $\mathcal{L}$ of Eq. (\ref{mod}):   
\begin{align}
\label{lag}
\mathcal{L}=&\frac{i}{2}\bigg( \psi^*\frac{\partial\psi}{\partial t}-\psi\frac{\partial\psi^*}{\partial t} \bigg)-|\psi|^2\nonumber\\
&\times\bigg[U\delta_T+\frac{1}{2}\int d\mathbf{r} 'K(\mathbf{r}-\mathbf{r}')|\psi(\mathbf{r}')|^2\bigg]+|\nabla\psi|^2,
\end{align}
and study the ground state for $U=0$ by the following Gaussian ansatz: $\psi=\sqrt{\frac{2P}{\pi}}\frac{e^{-r^2/a^2}}{a}$, defined by the power $P=\langle\psi|\psi\rangle$ and waist $a(t)$. By substituting the ansatz in (\ref{lag}), after performing a variational derivative over $a$, we obtain a classical dynamics described by the following Hamiltonian $\mathcal{H}$:
\begin{align}
\label{gs}
&\mathcal{H}=\frac{1}{2}\bigg(\frac{\partial a}{\partial t}\bigg)^2+\mathcal{V}, &\mathcal{V}=\frac{8}{a^2}-\frac{P}{2\pi\sigma^2}Z(a),
\end{align} 
with $Z(x)=e^x\Gamma(0,-x)$ and $\mathcal{V}$ acting as a potential for the one dimensional motion of $a$. The potential $\mathcal{V}$ has a bell shape profile that possesses a unique absolute minimum $V(a^*)$ for every combination of $P$ and $\sigma$. The fixed point $a(0)=a^*$ corresponds to a soliton wave of the system, which propagates in a translational fashion with fixed waist $a(t)=a^*$, while different initial values lead to a breather \cite{Snyder97} characterized by a periodic oscillation of $a$ in time. When the kicks are turned on, for $U\neq 0$, the dynamics of the ground state is perturbed by an addition of momentum $\mathbf{p}=(p_x,p_y)$, with a consequent translation of its center of mass. In order to model this dynamics, we considered the following general ansatz for the ground state evolution: $\psi=\sqrt{\frac{2P}{\pi}}\frac{e^{-(\mathbf{r}-\mathbf{r_0})^2/a^2+i\mathbf{p}(\mathbf{r}-\mathbf{r_0})/2T}}{a}$, with $\mathbf{p}(t)$, $a(t)$ and $\mathbf{r}_0(t)=[x_0(t),y_0(t)]$ Lagrangian variables whose equations of motion, after an integration from $nT$ to $(n+1)T$, are found to be:
\begin{align}
\label{sm0}
&\mathbf{p}_{n+1}=\mathbf{p}_n-[\gamma_n\mathbf{g}+\epsilon_n\mathbf{u}\sin(x_0+y_0)],\nonumber\\
&\mathbf{r}_{n+1}=\mathbf{r}_n+\mathbf{p}_{n+1},
\end{align}
with $\gamma_n=K e^{-\frac{a_n^2}{8}}$, $\epsilon_n=2\epsilon T e^{-\frac{a_n^2}{4}}$,  $\mathbf{g}=[\sin x_0,\sin y_0]$, $\mathbf{u}=[1,1]$, $f_n\equiv f(nT)$ and $a_{n+1}$ calculated from the integration of the following equation:
\begin{align}
\label{wa1}
\frac{\partial^2 a}{\partial t^2}=&-\frac{\partial\mathcal{V}}{\partial a}+\delta_Te^{-\frac{a^2}{8}}\nonumber\\
&\times[\gamma(\cos x_0+\cos y_0)+2e^{-\frac{a^2}{8}}\epsilon\cos(x_0+y_0)].
\end{align}
Equations (\ref{sm0}) can be regarded as a variant of the four dimensional standard map, which is randomized by time dependent coupling parameters $\gamma_n$ and $\epsilon_n$. The latter depend on Eq. (\ref{wa1}), which represents the motion of a two dimensional nonlinear kicked rotor. The system possesses an $a$ dependent chaos parameter, given by $K_a=e^{-a^2/8}\gamma T$. For $K>K^*$, Eq. (\ref{wa1}) is fully chaotic and can be regarded as an external noise source to Eqs. (\ref{sm0}), increasing the mixing of the overall system \cite{PhysRevLett.61.655}. To highlight such a dynamics, we plot in Fig. \ref{lyap}a and Fig. \ref{lyap}b the positive Lyapunov exponent $\lambda$ \cite{OTT} calculated for Eqs. (\ref{sm0}) and Eq. (\ref{wa1}), respectively. As seen in Fig. \ref{lyap}a, Eqs. (\ref{sm0}) show a strong hyperchaotic behavior, with two positive Lyapunov exponents whose largest value grows linearly with $\epsilon$. Fig. \ref{lyap}b, conversely, displays the chaotic nature of wave packet extension $a$, whose Lyapunov coefficient $\lambda$ increases significantly fast (quadratically) with coupling.
\begin{figure}
\centering
\includegraphics[width=8.5cm]{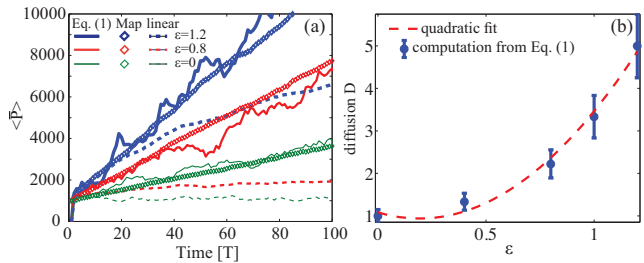}
\caption{
\label{diff}
(Color Online). (a) Momentum diffusion $\langle\bar P\rangle$ versus time calculated from Eq. (\ref{mod}) (solid lines), Eqs. (\ref{sm0})-(\ref{wa1}) (diamond markers) and Eq. (\ref{mod}) in linear regime (dashed lines); (b) diffusion coefficient $D$ versus coupling $\epsilon$. In the simulations we set $\sigma=0.2$ and $K=5$.}
\end{figure}
We investigate the diffusion in momentum $\mathbf{p}$ by observing that above the stochastization threshold $K>K^*$, the change in momentum $\Delta\mathbf{ p}=\mathbf{p}_{n+1}-\mathbf{p}_n\propto K$ becomes large compared to $\pi$. The classic position $\mathbf{r}_n$, which is taken modulo $2\pi$, can be treated a random process, statistically uncorrelated in time and uniformly distributed in $[-\pi,\pi]$. The diffusion coefficient $D$ is therefore evaluated as follows:
\begin{align}
\label{deq}
&D=\bigg\langle\frac{\Delta\mathbf{p}_n^2}{2}\bigg\rangle=\frac{K^2}{2}\langle \sin x_0^2\rangle\langle e^{-a_n^2/8}\rangle+2\epsilon^2 T^2\langle e^{-a_n^2/4}\rangle\nonumber\\
&\times \langle \sin (x_0+y_0)^2\rangle=\frac{K^2}{4}\langle e^{-a_n^2/8}\rangle+\epsilon^2 T^2\langle e^{-a_n^2/4}\rangle,
\end{align}
To evaluate the averages $\langle e^{-a_n^2/8}\rangle$ and $\langle e^{-a_n^2/4}\rangle$, we can consider $a$ as a random variable (due to its chaotic motion in the phase-space), uniformly distributed between its oscillation extrema $a_{min}$ and $a_{max}$:
\begin{align}
\label{expa}
\langle &e^{-a_n^2/\tau^2}\rangle=\frac{\sqrt{\pi}\tau}{2\Delta}\nonumber\\
&\times\bigg[\mathrm{erf}\bigg(\frac{a_{max}}{\tau}\bigg)-\mathrm{erf}\bigg(\frac{a_{min}}{\tau}\bigg)\bigg]=1+O(a_{max}^3/\tau^3)
\end{align}
being $\Delta=a_{max}-a_{min}$ and having expanded the error functions up to second order, due to the smallness of their arguments $a/\tau < 1$. By substituting (\ref{expa}) into (\ref{deq}), we obtain the diffusion coefficient, which reads as follows:
\begin{equation}
\label{ddd}
D=\frac{K^2}{4}+\epsilon^2 T^2
\end{equation}
Equation (\ref{ddd}) allows to derive interesting properties for the nonlinear dynamics of Eq. (\ref{mod}). In particular, the quantum average $\langle P\rangle$ results from an hyperchaotic system described by a four dimensional standard map with random coefficients, and each realization manifests itself as a random walk in Fig. \ref{sim}b. The map diffusion rate is identical to the momentum diffusion of the classical linear rotor \cite{PhysRevLett.61.659}, hence, an additional average (in time or over an ensemble of input conditions) re-establishes a perfect classical correspondence for every coupling $\epsilon\ge 0$. It is worthwhile observing that the classical correspondence in the multidimensional linear quantum rotor is manifested only for very high coupling $\epsilon$, and in general the quantum diffusion $\langle P\rangle$ follows a fractional behavior with $\langle P\rangle\propto t^{\beta<1}$ (see e.g., \cite{PhysRevLett.61.659} or Fig. \ref{sim}b dashed lines). As a result, the linear quantum rotor sub-diffuses at a slower rate than its classical counterpart. Conversely, Eq. (\ref{ddd}) predicts a perfect classical correspondence for every coupling $\epsilon$, which is re-established thanks to nonlinear effects. In order to demonstrate this dynamics, we performed extensive numerical simulations from Eq. (\ref{mod}) and calculated the average diffusion through a quantum average followed by an average over different input conditions
$\langle\bar P\rangle=\int D\psi\big\langle\psi\big |\frac{\hat{p}^2}{2}\big|\psi\big\rangle$.
Figure \ref{diff}a summarizes our results obtained for $K=5$, $\sigma=0.2$ and by considering an initial wave packet composed by a Gaussian beam with waist $\omega_0=0.3$ and amplitude $A=4$. In complete agreement with Eqs. (\ref{sm0})-(\ref{ddd}), we observe a diffusive behavior $\langle\bar P \rangle\propto t$ for every $\epsilon\ge 0$ (Fig. \ref{diff}a solid lines and diamond markers), whose rate is significantly faster than the linear subdiffusive dynamics (Fig. \ref{diff}a dashed lines). We can therefore conclude that nonlinearity favors the energy transport in the system, increasing diffusion through nonlinear wave-particles that are faster than their linear counterparts. This result also highlights the intimate connection between the wave-particle aspects of nonlinear waves, whose quantum-classical characters cannot be singularly broken, but conversely emerge naturally after averaging over the corresponding degree of freedom. To further verify the scaling dependence predicted by Eq. (\ref{ddd}), we calculated the diffusion coefficient $D$ of the nonlinear system for increasing $\epsilon$ (Fig. \ref{diff}b). In perfect agreement with our theory, we observe a quadratic behavior versus the coupling parameter $\epsilon$.\\ 
In conclusion, motivated by the large interest in the study of energy transport in complex media, we investigated the quantum-classical correspondences in many-dimensional quantum chaos. In particular, we employed a two-dimensional, nonlinear quantum kicked rotor (NQKR) and study the role of nonlinearity in enhancing or depleting energy diffusion and quantum-classical correspondences. We analytically tackled the problem by a variational analysis, reducing the dynamics to a four-dimensional standard map with random coefficients. In such an hyperchaotic system, a perfect classical correspondence is established by nonlinearity and an enhanced diffusion is observed due to solitons wave-particles, which are able to diffuse energy at a faster rate with respect to linear waves. These results show that quantum time reversal of classical irreversible systems is completely prevented in many dimensions, and demonstrate that nonlinearity can be effectively employed to increase the transport of energy in complex media. This work is expected to stimulate further theory and experiments in the broad area dealing with quantum chaos and energy transport phenomena. \\
A. Fratalocchi thanks S. Trillo for fruitful discussions. We acknowledge funding from KAUST (Award No. CRG-1-2012-
FRA-005).

%

\end{document}